\documentclass[cameraready]{Interspeech}
\usepackage{booktabs}
\usepackage{multirow}
\usepackage{makecell}
\usepackage{xcolor}
\usepackage{graphicx}
\usepackage{algorithm}
\usepackage{subcaption} % 必须引入这个包
\usepackage{algpseudocode}
\usepackage[algo2e]{algorithm2e}
\newcommand{\low}[1]{\underline{#1}}

\title{DuraMark: Duration-Embedded Watermarking in LLM-based TTS}

\author[affiliation={1}]{Zhenwei}{Mou}
\author[affiliation={2}]{Weili}{Jiang}
\author[affiliation={1}]{Liping}{Chen}
\author[affiliation={1}]{Zhen-Hua}{Ling}
\author[affiliation={3}]{Kong Aik}{Lee}
\author[affiliation={2}]{Kai}{Gao}
\author[affiliation={2}]{Boyu}{Zhao}
\address{
    $^1$ University of Science and Technology of China, China \\
    $^2$ Institute of Forensic Science, Ministry of Public Security, China \\
    $^3$ The Hong Kong Polytechnic University, China
}
\email{zwmu@mail.ustc.edu.cn, \{lipchen, zhling\}@ustc.edu.cn, \{jiangweili, gaokai\}@cifs.gov.cn, zhaoboyu\_uestc@163.com, kong-aik.lee@polyu.edu.hk \thanks{\emph{Corresponding author: Liping Chen.}} \thanks{This work was supported in part by the National Key Research and Development Program Project 2024YFE0217200, the Innovation and Technology Fund of the Hong Kong SAR MHP/048/24, and the National Natural Science Foundation of China under Grant 62506349 and U23B2053.}}


\usepackage{comment}

\begin{document}

\maketitle

% the abstract here must exactly match the abstract entered into the paper submission system
\begin{abstract}
    % 1000 characters. ASCII characters only. No citations.
    Large language model (LLM)-based text-to-speech (TTS) models have achieved remarkable voice cloning capabilities, raising concerns about potential deepfake misuse. Speech watermarking mitigates this by embedding traceable information into generated speech. Mainstream watermarking methods operate at the signal level (waveform or spectrogram), rendering the watermark vulnerable to generative attacks (e.g., neural codec and vocoder). To address this, we propose DuraMark, a robust information-level watermarking framework. It utilizes syllable duration editing to achieve watermark embedding. Specifically, DuraMark integrates a duration-controllable LLM-based TTS model to edit syllable durations during synthesis, coupled with a duration extractor to extract these durations for detection. Experiments demonstrate DuraMark's superior robustness against generative attacks, significantly outperforming signal-level baselines. Audio samples are available at \url{https://muzw.github.io/duramark_demo/}.
\end{abstract}

\vspace{-0.6em}
\section{Introduction}
\vspace{-0.2em}

Large language model (LLM)-based text-to-speech (TTS) models \cite{valle,cosyvoice,fireredtts} have achieved remarkable progress in synthesizing speech with high fidelity and speaker similarity. Such advancements inevitably raise concerns regarding potential deepfake misuse. Speech watermarking mitigates these risks by embedding imperceptible watermarks into synthesized speech, effectively distinguishing AI-generated content from natural speech.

Currently, mainstream speech watermarking methods operate at the signal level. Traditional methods \cite{lsb,spread_spectrum,qim} operate in the time or frequency domains, relying on domain-specific features to design the watermark embedding and decoding functions. Neural network (NN)-based methods have emerged in two main categories: post-processing and generative watermarking. Post-processing methods like AudioSeal \cite{audioseal}, WavMark \cite{wavmark}, and MaskMark \cite{maskmark} employ an embedder-detector architecture. The embedder performs signal-level reconstruction, taking both the watermark and the speech as input to synthesize the watermarked waveform or spectrogram. Subsequently, the detector recovers the watermark from the reconstructed signal. Alternatively, some generative watermarking methods \cite{hifiganw,traceablespeech} employ neural vocoder or codec architecture as the embedder. By replacing these components within generative models, the watermarking mechanism is integrated into the speech synthesis process. Beyond signal-level approaches, information-level watermarking was explored in \cite{pitch_watermark}, which embeds watermarks by editing pitch values of speech segments.

Robustness against attacks is a critical metric for watermark models. Signal-level watermarks are vulnerable to generative attacks such as NN-based resynthesis, including neural audio codecs \cite{encodec,dac,speechtokenizer} and vocoders \cite{bigvgan,vocos,hifigan}. Generative attacks reconstruct speech at the information level such as content, semantics, and prosody. The signal-level watermark is smoothed as redundant signal details, leading to a reduction of the watermark information. The information-level method \cite{pitch_watermark} relies on traditional signal post-processing to edit pitch values, inevitably resulting in unnatural prosody. To address these issues, we propose DuraMark, a novel generative watermarking framework operating at the information level. Specifically, our investigation is conducted within a duration-controllable LLM-based TTS framework, coupled with a duration extractor. Watermarks are embedded by editing syllable durations (in frames) using the duration-controllable TTS model. The detection phase employs the duration extractor to extract the duration sequence from speech and compare it against the target watermark for detection. Our main contributions are summarized as follows:
\begin{itemize}
\item We develop a duration-controllable LLM-based TTS model, allowing for precise syllable-level duration editing.
\item We introduce DuraMark, a novel generative watermarking framework based on the duration-controllable LLM-based TTS model. Experiments demonstrated that DuraMark achieved superior robustness, particularly against NN-based resynthesis, significantly outperforming the existing signal-level methods.
\end{itemize}

\begin{figure*}[h]
     \centering
     \vspace{-1.5em}
     % 第一张图：较窄
     \begin{subfigure}[b]{0.65\textwidth}
         \centering
         \includegraphics[width=\textwidth]{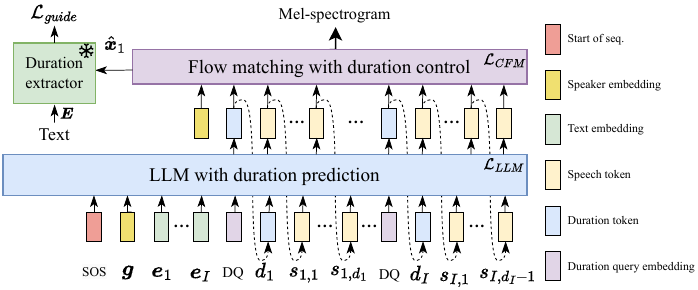}
         \caption{Duration-controllable TTS}
         \vspace{-0.5em}
         \label{fig:duration_control_tts}
     \end{subfigure}
     \hfill % 自动填充间距，保证左右对齐
     % 第二张图：较宽
     \begin{subfigure}[b]{0.32\textwidth}
         \centering
         \includegraphics[width=\textwidth]{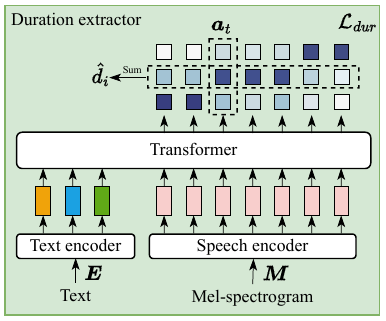}
         \caption{Duration extractor}
         \vspace{-0.5em}
         \label{fig:extractor}
     \end{subfigure}

     \caption{Architectures of the duration-controllable TTS and duration extractor used in DuraMark.}
     \vspace{-1.5em}
     \label{fig:duramark}
\end{figure*}

%\section{Architecture}
\vspace{-0.6em}
\section{Duration-controllable LLM-based TTS}
\vspace{-0.2em}

Unlike conventional LLM-based TTS that models duration implicitly, the proposed method introduces explicit duration control and precise duration extraction. As illustrated in Fig. \ref{fig:duramark} (\subref{fig:duration_control_tts}), the duration-controllable TTS integrates an LLM with a flow matching decoder. Given the syllable-level text embedding sequence $\boldsymbol{E}=\{\boldsymbol{e}_1, \dots, \boldsymbol{e}_I\}$ as input with $I$ denoting the number of syllables, the LLM predicts duration tokens $\boldsymbol{d} = \{d_1, \dots, d_I\}$ and speech tokens $\boldsymbol{S} = \{\boldsymbol{s}_1, \dots, \boldsymbol{s}_I\}$, where
$\boldsymbol{s}_i = \{s_{i,1},\dots, s_{i,d_i}
\}$ contains $d_i$ frames within the $i$-th syllable. The flow matching decoder then synthesizes Mel-spectrograms conditioned on $\boldsymbol{d}$ and $\boldsymbol{S}$. During training, the LLM and flow matching decoder are trained independently. Specifically, the flow matching decoder employs a pre-trained duration extractor to provide auxiliary guidance, ensuring that the synthesized speech strictly adheres to the provided durations, i.e., $\boldsymbol{d}$. During inference, to achieve explicit duration control, the durations predicted by the LLM can be edited and subsequently fed into the flow matching decoder to synthesize speech with the specified durations.

\vspace{-0.6em}
\subsection{LLM with duration prediction}
\vspace{-0.2em}
The LLM generation operates autoregressively syllable by syllable. For the $i$-th syllable, a duration query (\texttt{DQ}) embedding triggers the prediction of the duration token $d_i$, followed by $d_i$ speech tokens. Conditioned on the speaker embedding $\boldsymbol{g}$, text embedding $\boldsymbol{E}$, and generated history, the probability distribution of $d_i$ is formulated as:
\vspace{-0.2em}
\begin{equation}
\label{eq:duration_predict}
\boldsymbol{y}_i=p\left({\mathcal C}^{\rm d} \mid {\boldsymbol g}, \boldsymbol{E}, {d}_{1:i-1}, {{\boldsymbol s}}_{1:i-1}\right),
\end{equation}
where ${\mathcal C}^{\rm d}$ denotes the duration token set. The probability of the $t$-th frame token within $i$-th syllable $s_{i,t}$ is predicted with:
\vspace{-0.2em}
\begin{equation}
\boldsymbol{z}_{i,t} = p\left(\mathcal{C}^{\rm s} \mid {\boldsymbol g}, \boldsymbol{E},  {d}_{1:i}, {{\boldsymbol s}}_{1:i-1}, s_{i,1:t-1}\right),
\end{equation}
where $t = 1,\dots,d_{i}$ and ${\mathcal C}^{\rm s}$ is the speech token set. Let $\tilde{\boldsymbol{y}}_i$ and $\tilde{\boldsymbol{z}}_{i,t}$ denote the ground-truth distributions for duration and speech tokens, respectively. The training objective for the LLM is defined by the cross-entropy loss:
\vspace{-0.2em}
\begin{equation}
\begin{split}
\mathcal{L}_{\text{llm}} & = -w_{\text{llm}}\frac{1}{I}\sum_{i=1}^{I}\tilde{\boldsymbol{y}}_i\log \boldsymbol{y}_i \\
& \quad -(1-w_{\text{llm}})\frac{1}{\sum_{i=1}^{I} d_i}\sum_{i=1}^{I}\sum_{t=1}^{d_i}\tilde{\boldsymbol{z}}_{i,t}\log \boldsymbol{z}_{i,t},
\end{split}
\end{equation}

where $0\le w_{\text{llm}}\le 1$ is the weight variable.

During inference, $d_i$ and $s_{i,t}$ are sampled from $\boldsymbol{y}_i$ and $\boldsymbol{z}_{i,t}$, respectively.

\vspace{-0.2em}
\subsection{Duration extractor}
\label{ssec:extractor}
\vspace{-0.2em}
The duration extractor takes the syllable sequence $\boldsymbol{E}$ and the Mel-spectrogram $\boldsymbol{M} = \{\boldsymbol{m}_1, \dots, \boldsymbol{m}_T\}$ as inputs to extract the duration for each syllable. As illustrated in Fig. \ref{fig:duramark}(\subref{fig:extractor}), the encoded text and Mel-spectrogram features are concatenated and fed into a transformer. For each frame $\boldsymbol{m}_t$, the model predicts a probability distribution over the input syllables, denoted as $\boldsymbol{a}_t = \{a_{1,t}, \dots, a_{I,t}\}$, where $a_{i,t}$ represents the probability that the $t$-th frame belongs to the $i$-th syllable.

Let $\tilde{\boldsymbol{a}}_t$ denote the ground-truth distribution for the $t$-th frame. The extractor's training objective is to minimize frame-level cross-entropy loss:
\vspace{-0.2em}
\begin{equation}
\label{eq:dur_loss}
\mathcal{L}_{\text{dur}}(\tilde{\boldsymbol{a}}_t,\boldsymbol{a}_t)= -\frac{1}{T}\sum_{t=1}^{T} \tilde{\boldsymbol{a}}_t \log \boldsymbol{a}_t.
\end{equation}
\vspace{-0.2em}

During inference, the estimated duration $\hat{d}_i$ for the $i$-th syllable is obtained by summing its corresponding frame probabilities:
\vspace{-0.2em}
\begin{equation}
\label{eq:estimate_duration}
\hat{d}_i = \sum_{t=1}^T a_{i,t}.
\end{equation}
\vspace{-0.2em}
\vspace{-1em}
\subsection{Flow matching with duration control}
\label{sssec:decoder}
\vspace{-0.2em}
The decoder employs Optimal Transport Conditional Flow Matching (OT-CFM) \cite{flow} to synthesize Mel-spectrograms. OT-CFM defines a linear probability flow $\psi_t(\boldsymbol{x}_0) = (1 - t)\boldsymbol{x}_0 + t\boldsymbol{x}_1$ that transitions from Gaussian noise $\boldsymbol{x}_0$ to the target Mel-spectrogram $\boldsymbol{x}_1$.  The target vector field is $\boldsymbol{u}_t = \boldsymbol{x}_1 - \boldsymbol{x}_0$. The model $\boldsymbol{v}_t$ is trained to regress this field conditioned on speaker embedding, speech tokens and duration tokens $\boldsymbol{c} = \{ \boldsymbol{g}, \boldsymbol{S}, \boldsymbol{d}\}$ by minimizing:
\vspace{-0.2em}
\begin{equation}
\mathcal{L}_{\text{CFM}} = \mathbb{E}_{t, \boldsymbol{x}_0, \boldsymbol{x}_1} \left[ \lVert \boldsymbol{v}_t(\psi_t(\boldsymbol{x}_0), t, \boldsymbol{c}) - (\boldsymbol{x}_1 - \boldsymbol{x}_0) \rVert^2 \right].
\end{equation}

To achieve explicit duration control, the flow matching decoder utilizes a frozen duration extractor to guide the training. Specifically, the Mel-spectrogram $\hat{\boldsymbol{x}}_1$ is estimated via a single-step approximation:
\vspace{-0.2em}
\begin{equation}
\hat{\boldsymbol{x}}_1 = \psi_t(\boldsymbol{x}_0) + (1 - t) \cdot \boldsymbol{v}_t(\psi_t(\boldsymbol{x}_0), t, \boldsymbol{c}).
\end{equation}
This estimated Mel-spectrogram $\hat{\boldsymbol{x}}_1$ is then fed into the frozen duration extractor to obtain the predicted probability distribution $\hat{\boldsymbol{a}}_t$. The guidance loss is then computed between $\hat{\boldsymbol{a}}_t$ and ground-truth $\tilde{\boldsymbol{a}}_t$:
\vspace{-0.2em}
\begin{equation}
\mathcal{L}_{\text{guide}} = \mathcal{L}_{\text{dur}}(\tilde{\boldsymbol{a}}_t,\hat{\boldsymbol{a}}_t).
\end{equation}

The final training objective of flow matching decoder is a weighted summation:
\vspace{-0.2em}
\begin{equation}
\mathcal{L}_{\text{flow}} = w_{\text{flow}}\mathcal{L}_{\text{CFM}} + (1-w_{\text{flow}}) \mathcal{L}_{\text{guide}},
\end{equation} 
where $0\le w_{\text{flow}}\le 1$ is the weight variable.

\vspace{-0.6em}
\section{Duration-embedded watermarking}
\label{ssec:watermarking}
\vspace{-0.2em}

Given an input syllable sequence $\boldsymbol{E}=\{\boldsymbol{e}_1,\dots,\boldsymbol{e}_I\}$, the original duration for each syllable is first predicted by the duration-controllable TTS model in Fig. \ref{fig:duramark} (\subref{fig:duration_control_tts}). Each duration resides in an even state (representing bit `0') or an odd state (representing bit `1'), respectively. To perform embedding, a watermark sequence $\boldsymbol{w}=\{w_1,\dots,w_I\}$ of the same length as the syllable sequence is defined. Specifically, for the $i$-th syllable, $w_i \in \{0, 1\}$ dictates that its duration value should be even ($w_i=0$) or odd ($w_i=1$). The original durations are then edited to meet the target states. Thereby, the speech is generated using the edited durations, wherein the watermark is encoded into the duration sequence of the syllables. In detection, the correlation between the extracted syllable duration sequence and the watermark sequence is computed and compared with a threshold to reach a decision. 

\vspace{-0.6em}
\subsection{Embedding based on duration editing}
\vspace{-0.2em}

The embedding process is illustrated in phase 1 of Algorithm \ref{alg:watermark}. The LLM in Fig. \ref{fig:duramark} (\subref{fig:duration_control_tts}) takes the syllable sequence $\boldsymbol{E}$ and speaker embedding $\boldsymbol{g}$ as inputs. For the $i$-th syllable, the LLM first predicts the duration probability distribution $\boldsymbol{y}_i$ according to Eq. (\ref{eq:duration_predict}). Here, $\boldsymbol{y}_i$ is a vector of size $|{\mathcal C}^{\rm d}|$, wherein each element represents the probability of a specific duration (i.e., frame count). After sampling an initial duration $d_i$ from $\boldsymbol{y}_i$, it is explicitly edited to yield $d^*_i$. Specifically, if $d_i$ matches the target state (i.e., $d_i \bmod 2 = w_i$), the edited duration is kept as $d^*_i = d_i$. Otherwise, $d^*_i$ is set to either $d_i+1$ or $d_i-1$, depending on which candidate duration achieves a higher probability in $\boldsymbol{y}_i$. Conditioned on the edited duration $d^*_i$, the LLM subsequently predicts the speech tokens $\boldsymbol{s}_i$. Finally, the edited durations $\boldsymbol{d}^*=\{d^*_1,\dots,d^*_I\}$ and speech tokens $\boldsymbol{S} = \{\boldsymbol{s}_1, \dots, \boldsymbol{s}_I\}$ are fed into the flow matching decoder to synthesize speech $\hat{\boldsymbol{O}}$.

\vspace{-0.6em}
\subsection{Detection based on duration extraction}
\vspace{-0.2em}

The detection process follows phase 2 of Algorithm \ref{alg:watermark}. Given a speech utterance $\boldsymbol{O}$ and its corresponding transcription, the syllable sequence $\boldsymbol{E}$ is first obtained. Subsequently, the estimated duration sequence $\hat{\boldsymbol{d}}=\{\hat{d}_1, \dots, \hat{d}_I\}$ is derived via the extractor in Fig. \ref{fig:duramark}(b) according to Eq. (\ref{eq:estimate_duration}). To determine the presence of a watermark, $\hat{\boldsymbol{d}}$ is compared with the embedding $\boldsymbol{w}$. Since $\hat{\boldsymbol{d}}$ and $\boldsymbol{w}$ reside in different value domains, the continuous interval of $[-1, 1]$ is empirically used as the common target space for their comparison. Here, values closer to \(-1\) indicate a higher probability of being in an even state, while values closer to \(1\) indicate a higher probability of being in an odd state. The $i$-th elements of $\boldsymbol{w}$ and $\hat{\boldsymbol{d}}$ are mapped respectively as follows:
\begin{equation}
w'_i = 2w_i - 1,
\end{equation}
\begin{equation}
d'_i = -\cos(\pi \hat{d}_i).
\end{equation} 
A similarity score $\mathcal{T}$ is then calculated as the correlation between the mapped sequences $\boldsymbol{d}'$ and $\boldsymbol{w}'$:
\begin{equation}
\mathcal{T} = \frac{1}{I}\sum_{i=1}^I d'_i \cdot w'_i.
\end{equation}
Finally, the watermark is deemed present if the similarity score $\mathcal{T}$ exceeds a pre-defined threshold $\tau$.

\begin{algorithm}[t]
\footnotesize
    \caption{Duration-embedded watermarking pipeline.}
    \label{alg:watermark}

    \noindent\rule{\linewidth}{0.4pt}\par
    \textbf{Phase 1 - Embedding}\par
    \vspace{-4pt}
    \noindent\rule{\linewidth}{0.4pt}\par
    \noindent\textbf{Input:} syllable sequence $\boldsymbol{E}$, watermark sequence $\boldsymbol{w}=\{w_1,\dots,w_I\}$, speaker embedding $\boldsymbol{g}$ \\
    \noindent\textbf{Output:} watermarked speech \par
    \vspace{2pt}
    \expandafter{\romannumeral 1}. Generate duration tokens and speech tokens via the LLM in Fig. \ref{fig:duramark} (\subref{fig:duration_control_tts}) with duration editing:\\
    \textbf{Initialize}: $d^*_{1:0} = \emptyset$, $\boldsymbol{s}_{1:0} = \emptyset$. 

    \For{$i=1,\dots,I$}
    {
        Predict duration distribution $\boldsymbol{y}_i$: \\
        \hspace*{2em} $\boldsymbol{y}_i = p({\mathcal C}^{\rm d} \mid {\boldsymbol g}, \boldsymbol{E}, {d}^*_{1:i-1}, {{\boldsymbol s}}_{1:i-1})$. \\
        Sample duration $d_i$ from $\boldsymbol{y}_i$. \\
        \eIf{$d_i \bmod 2 == w_i$}
        {
            $d^*_i = d_i$.
        }
        {
            $d^*_i = \operatorname*{argmax}_{x \in \{d_i+1, d_i-1\}} p_{\boldsymbol{y}_i}(x)$.
        }
        Generate $d^*_i$ speech tokens $\boldsymbol{s}_i$.
    }
    \expandafter{\romannumeral 2}. Synthesize watermarked speech $\hat{\boldsymbol{O}}$ using the flow matching decoder, conditioning on $\boldsymbol{S} = \{\boldsymbol{s}_1, \dots, \boldsymbol{s}_I\}$ and $\boldsymbol{d}^*=\{d^*_1,\dots,d^*_I\}$.

    \noindent\rule{\linewidth}{0.4pt}\par
    \textbf{Phase 2 - Detection}\par
    \vspace{-4pt}
    \noindent\rule{\linewidth}{0.4pt}\par
    \noindent\textbf{Input:} speech $\boldsymbol{O}$, syllable sequence $\boldsymbol{E}$, watermark sequence $\boldsymbol{w}=\{w_1,\dots,w_I\}$, threshold $\tau$ \\
    \noindent\textbf{Output:} detection decision \par
    \vspace{2pt}
    \expandafter{\romannumeral 1}. Extract estimated durations $\hat{\boldsymbol{d}}=\{\hat{d}_1,\dots,\hat{d}_I\}$ via the duration extractor given $\boldsymbol{O}$ and $\boldsymbol{E}$ according to Eq. (\ref{eq:estimate_duration}). \\
    \expandafter{\romannumeral 2}. Map to continuous interval [-1, 1]: \\
    \hspace*{4em} $w'_i = 2w_i - 1$, \\
    \hspace*{4em} $d'_i = -\cos(\pi \hat{d}_i)$. \\
    \expandafter{\romannumeral 3}. Compute similarity score: \\
    \hspace*{4em} $\mathcal{T} = \frac{1}{I}\sum_{i=1}^I d'_i \cdot w'_i$. \\
    \expandafter{\romannumeral 4}. Decision: watermarked if $\mathcal{T} > \tau$, otherwise unwatermarked. \par
    
\end{algorithm}

\vspace{-0.6em}
\section{Experiments}
\vspace{-0.2em}
\vspace{-0.2em}
\subsection{Experimental setup}
\vspace{-0.2em}
Experiments were conducted on Mandarin Chinese, in which a character corresponds to a single syllable. The training data was derived from WenetSpeech \cite{zhang2022wenetspeech}, which contains 10,000 hours of speech. Syllable boundaries were determined using the Montreal Forced Aligner (MFA) \cite{mcauliffe2017montreal}. Evaluation was conducted on the test set of the AISHELL-3 dataset \cite{shi2020aishell}, which comprises recordings from 214 speakers. Our models employed the CosyVoice framework using its open-source implementation. The models were trained using the Adam optimizer with a learning rate of $1 \times 10^{-5}$ on eight MLU 580 GPUs. The loss balancing weights were set to $\lambda_{llm}=1$ and $\lambda_{flow}=4$. During inference, we applied top-$p$ ($p=0.8$) and top-$k$ ($k=25$) sampling for speech tokens, while employing greedy sampling for duration tokens.

\begin{table}[t]
    \centering
    \fontsize{8pt}{9pt}\selectfont % 精确设置字体为8pt，行距9.6pt
    \vspace{-0.5em}
    \caption{Impact of speech length on detection performance, measured by TPR.}
    \vspace{-1em}
    \label{tab:length_impact}
    \begin{tabular}{lcc}
        \toprule
        \textbf{Length (Syllables)} & \textbf{DuraMark-Info} & \textbf{DuraMark-Blind} \\
        \midrule
        17--32   & 0.981 & 0.942 \\
        33--64   & 0.998 & 0.987 \\
        65--100  & 0.998 & 0.992 \\
        \bottomrule
    \end{tabular}%
    \vspace{-0.3cm}
\end{table}

\begin{table*}[t]
\centering
\fontsize{8pt}{9.2pt}\selectfont % 精确设置字体为8pt，行距9.6pt
\renewcommand{\arraystretch}{1}
% 调整表格内的列间距，防止过宽
\setlength{\tabcolsep}{4pt}
\vspace{-2em}
\caption{Robustness comparison against baselines measured by TPR on the test set. Underlined values indicate TPR $< 0.90$.}
\label{tab:robustness_comparison}
\vspace{-1em}

% 使用 resizebox 自动缩放表格以适应文本宽度
\resizebox{0.98\textwidth}{!}{%
\begin{tabular}{l|c|c|c|c|c|c|c}
\toprule
\textbf{Attack Category} & \textbf{Perturbation} & \textbf{Param} & \textbf{AudioSeal}~\cite{audioseal} & \textbf{Timbre}~\cite{timbre} & \textbf{WavMark}~\cite{wavmark} & \textbf{DuraMark-Info} & \textbf{DuraMark-Blind}  \\
\midrule
\midrule
\textbf{None} & Original & - & 1.000 & 1.000  & 1.000 & 0.998 & 0.987 \\
\midrule
\multirow{7}{*}{\makecell[l]{\textbf{Neural}\\ \textbf{Codecs}}}
 & \multirow{2}{*}{EnCodec} & 6.0k & \low{0.773} & \low{0.124} & \low{0.015} & 0.991 & 0.968 \\
 & & 12.0k & 1.000 & \low{0.551} & \low{0.010} & 0.994 & 0.970 \\
 \cline{2-8}
 & \multirow{2}{*}{DAC} & 3.0k & \low{0.096} & \low{0.414} & \low{0.007} & 0.994 & 0.980 \\
 & & 4.5k & \low{0.512} & \low{0.832} & \low{0.002} & 0.996 & 0.979 \\
 \cline{2-8}
 & SpeechTokenizer & 4.0k & \low{0.004} & \low{0.047} & \low{0.017} & 0.984 & 0.966  \\
 \cline{2-8}
 & FACodec & 2.4k & \low{0.013} & \low{0.036} & \low{0.010} & 0.994 & 0.977  \\
\midrule
\multirow{3}{*}{\makecell[l]{\textbf{Neural}\\ \textbf{Vocoders}}}
 & BigVGAN & - & 0.908 & 1.000  & \low{0.012} & 0.997 & 0.987  \\
 & Vocos & - & \low{0.005} & 1.000  & \low{0.010} & 0.997 & 0.979  \\
 & HiFiGAN & - & \low{0.013} & 1.000  & \low{0.007} & 0.994 & 0.985 \\
\midrule
\multirow{2}{*}{\makecell[l]{\textbf{Speech}\\ \textbf{Enhancement}}}
 & FRCRN & - & 1.000 & 1.000  & 1.000 & 0.999 & 0.989  \\
 & Demucs & - & 1.000 & 1.000  & 1.000 & 0.998 & 0.983 \\
\midrule
\multirow{3}{*}{\makecell[l]{\textbf{Lossy}\\ \textbf{Compression}}}
 & MP3 & 32k & 1.000 & 1.000  & 1.000 & 0.998 & 0.983  \\
 & Opus & 16k & 0.986 & 1.000  & \low{0.665} & 0.994 & 0.983  \\
 & Quantization & $2^6$ & 1.000 & 1.000  & \low{0.010} & 0.995 & 0.982  \\
\midrule
\multirow{4}{*}{\makecell[l]{\textbf{Signal}\\ \textbf{Processing}}}
 & Gaussian Noise & 20dB & 1.000 & 0.999  & \low{0.050} & 0.994 & 0.979  \\
 & Background Noise & 20dB & 1.000 & 1.000  & 0.905 & 0.987 & 0.965  \\
 & Low-pass & 4.8kHz & 1.000 & 1.000  & 1.000 & 0.973 & 0.961  \\
 & Smoothing & 18 & 1.000 & 1.000  & 0.945 & 0.995 & 0.978  \\
\midrule
\multicolumn{3}{c|}{\textbf{Average}} & 0.701 & 0.790 & 0.403 & \textbf{0.993} & 0.978 \\
\bottomrule
\end{tabular}%
}
\vspace{-0.5cm}
\end{table*}

\vspace{-0.6em}
\subsection{Baseline and proposed methods}
\vspace{-0.2em}
We compared DuraMark against three state-of-the-art signal-level baselines using their official implementations: (1) \textbf{AudioSeal} \cite{audioseal}, a localized time-domain method using a neural generator; (2) \textbf{Timbre} \cite{timbre}, a frequency-domain method manipulating the magnitude spectrum; and (3) \textbf{WavMark} \cite{wavmark}, an invertible spectrogram-based encoder-decoder framework. For the proposed method, we consider two detection scenarios: informed detection (\textbf{DuraMark-Info}) using ground-truth text, and blind detection (\textbf{DuraMark-Blind}) using text transcribed via Automatic Speech Recognition (ASR). In all experiments, the embedded watermark sequences were randomly generated.

\vspace{-0.6em}
\subsection{Impact of speech length}
\label{sec:impact}
\vspace{-0.2em}
Table \ref{tab:length_impact} analyzes the impact of speech length on detection performance. Test samples are categorized into three syllable-count intervals (17--32, 33--64, 65--100). For each configuration, 1,000 pairs of watermarked and unwatermarked utterances were generated. We evaluate performance using the True Positive Rate at 1.0\% False Positive Rate (TPR@1\%FPR). Results indicate that longer speech improves TPR due to the embedding of a longer watermark sequence. In subsequent experiments, we adopted the 33--64 syllable configuration.

\vspace{-0.6em}
\subsection{Robustness}
\vspace{-0.2em}
To comprehensively evaluate robustness, the following attacks are considered:
\textbf{1) Neural Codecs:} EnCodec \cite{encodec}, DAC \cite{dac}, SpeechTokenizer \cite{speechtokenizer}, and FACodec \cite{naturalspeech3};
\textbf{2) Neural Vocoders:} BigVGAN \cite{bigvgan}, Vocos \cite{vocos}, and HiFiGAN \cite{hifigan};
\textbf{3) Speech Enhancement:} Demucs \cite{demucs} and FRCRN \cite{frcrn};
\textbf{4) Lossy Compression:} MP3, Opus, and bit-depth quantization;
\textbf{5) Signal Processing:} Gaussian/background noise, low-pass filtering, and window smoothing. Baseline methods are applied to generated unwatermarked speech for a comparison with DuraMark. The results are summarized in Table \ref{tab:robustness_comparison}. In attack scenarios involving traditional signal processing, lossy compression, and neural speech enhancement, both the baselines and our method demonstrate robust performance. However, a significant performance gap is observed under neural audio codec and vocoder attacks. Baseline methods exhibit varying degrees of vulnerability; for instance, WavMark fails to resist most neural perturbations, while AudioSeal and Timbre show instability, performing well in some cases but dropping sharply in others. In contrast, DuraMark consistently maintains a TPR $> 95$\%, demonstrating superior robustness, particularly against generative attacks.

\begin{table}[h]
\centering
\fontsize{8pt}{9pt}\selectfont % 精确设置字体为8pt，行距9.6pt
\caption{Naturalness evaluation of the compared methods. The MOS results are reported with 95\% confidence intervals.}
\label{tab:naturalness}
\vspace{-1em}
\begin{tabular}{l|ccc}
\toprule
\textbf{Method} & \textbf{CER (\%)} $\downarrow$  & \textbf{MOS} $\uparrow$ \\
\midrule
Ground Truth & 5.44 & 4.35 $\pm$ 0.11 \\
Unwatermarked & 8.73 & 4.05 $\pm$ 0.09 \\
\midrule
AudioSeal \cite{audioseal} & 8.15 & 4.07 $\pm$ 0.08 \\
Timbre \cite{timbre} & 8.56 & 4.03 $\pm$ 0.07 \\
WavMark \cite{wavmark} & 9.25 & 3.97 $\pm$ 0.08 \\
DuraMark & 8.54 & 4.04 $\pm$ 0.07 \\
\bottomrule
\end{tabular}%
\vspace{-0.2cm}
\end{table}

\vspace{-0.6em}
\subsection{Naturalness}
\vspace{-0.2em}
We evaluate speech naturalness using both objective and subjective metrics. Specifically, intelligibility is measured by the Character Error Rate (CER) using Whisper \cite{radford2023robust}, while naturalness is evaluated via the Mean Opinion Score (MOS). For subjective evaluation, 20 sentences were randomly selected and assessed by eleven native speakers. As shown in Table \ref{tab:naturalness}, both the signal-level baselines and DuraMark impose minimal impact on CER and MOS compared with unwatermarked speech. This demonstrates that our method effectively preserves speech naturalness, even with duration-embedded watermarking.

\vspace{-0.6em}
\subsection{Ablation study}
\label{sec:ablation}
\vspace{-0.2em}
To validate the effectiveness of specific modules, we conducted ablation studies by: (1) removing duration tokens from the decoder input (w/o duration input); and (2) removing the duration guidance loss during training (w/o $\mathcal{L}_{\text{guide}}$). The ablation results are presented in Table \ref{tab:ablation_arch}. Removing either the duration input or the guidance loss leads to significant performance degradation. This indicates that without explicit duration input and guidance, the decoder fails to strictly adhere to the edited durations, resulting in a reduction of watermark information.

\begin{table}[t]
\centering
\fontsize{8pt}{9pt}\selectfont % 精确设置字体为8pt，行距9.6pt
\caption{Ablation results on TTS architecture measured by TPR.}
\label{tab:ablation_arch}
\vspace{-1em}
\begin{tabular}{lcc}
\toprule
\textbf{Method} & \textbf{Informed} & \textbf{Blind} \\
\midrule
\textbf{DuraMark} & \textbf{0.998} & \textbf{0.987} \\
\midrule
w/o duration input & 0.455 & 0.473 \\ 
w/o $\mathcal{L}_{\text{guide}}$ & 0.327 & 0.342 \\
\bottomrule
\end{tabular}%
\vspace{-0.5cm}
\end{table}

\vspace{-0.6em}
\section{Conclusion}
\label{sec:conclusion}
\vspace{-0.2em}
In this paper, we proposed DuraMark, a robust generative speech watermarking framework that embedded watermarks at the information level. Specifically, the watermark is embedded into the syllable duration sequence of generated speech. By first developing a duration-controllable LLM-based TTS model, DuraMark embedded watermark by editing syllable durations during the speech synthesis process. Meanwhile, a duration extractor is applied to extract the duration sequence from speech for watermark detection. Experimental results demonstrate that DuraMark significantly improves robustness, particularly against generative attacks, outperforming state-of-the-art signal-level methods. Furthermore, the naturalness evaluation confirms that the proposed method preserves high speech quality.

\clearpage

\bibliographystyle{IEEEtran}
\bibliography{refs}

\end{document}